\documentclass[10pt, aps, prb, amsmath,amssymb, twocolumn,
floatfix, showpacs, nofootinbib, longbibliography]{revtex4-2}
\usepackage[x11names, svgnames, dvipsnames]{xcolor}
\definecolor{maroon}{RGB}{128,0,0}
\usepackage{hyperref}
\hypersetup{colorlinks,linkcolor={maroon},citecolor={maroon},urlcolor={maroon}}
\usepackage{cleveref}
\usepackage{graphicx}
\usepackage{subfigure}
\usepackage{dcolumn}
\usepackage{bm}
\usepackage{comment}
\usepackage{braket}
\usepackage{textcomp, gensymb}
\crefname{figure}{Fig.}{Figures}
\crefname{section}{Sec.}{Secs.}

\begin{document}

\title{Buckling and flat bands in twisted bilayer graphene}

\author{Jannes van Poppelen}
\email{jannes.vanpoppelen@physics.uu.se}
\author{Annica M. Black-Schaffer}
\affiliation{Department of Physics and Astronomy, Uppsala University, Box 516, SE-75237 Uppsala, Sweden}%

\date{\today}

\begin{abstract}
Magic-angle twisted bilayer graphene (TBG) with its flat bands provides a rich platform for exploring emergent electronic orders. Similarly, periodically buckled monolayer graphene has been proposed as a tunable alternative for realizing flat bands. Here, we investigate the combined effect of buckling and twisting in bilayer graphene.
We find that periodic buckling in large-angle TBG initially enhances band flattening compared to monolayer graphene, but for sufficiently strong buckling, it instead increases the band dispersion. This occurs both because of the presence of interlayer coupling, which reduces the in-plane kinetic energy, and due to the opening of a gap at the Dirac point resulting from inversion-symmetry breaking. 
Additionally, we find that buckling-induced band flattening competes with twist-induced band flattening. While the former breaks sublattice symmetry, generating a sublattice polarization, the latter prefers to preserve it. This prevents buckling from generating even flatter bands at the magic angle. 
Nevertheless, we find that buckled TBG can exhibit flatter bands than pristine TBG over a wide range of twist angles, with a flatness similar to that of pristine magic-angle TBG.
\end{abstract}

\maketitle

\section{\label{sec:level1} Introduction}
Twisted bilayer graphene (TBG) is the prototypical moiré material. A slight rotational misalignment between two graphene layers has been shown to give rise to a rich variety of correlated phases, including the anomalous Hall effect \cite{tseng2022anomalous}, correlated insulating phases \cite{cao2018correlated}, unconventional superconductivity of still-uncertain origin \cite{cao2018unconventional, yankowitz2019tuning, oh2021evidence, lothman2022nematic}, strange-metal behavior \cite{cao2020strange}, and topological insulating phases \cite{nuckolls2020strongly}. It is the emergence of flat bands with a high density of states (DOS) at so-called “magic” twist angles \cite{suarez2010flat, bistritzer2011moire, tarnopolsky2019origin}, which strongly enhances the role of electronic interactions and thereby drives the emergence of the correlated phases. The electronic structure of TBG has also been shown to be controllable beyond the twist angle, with electrostatic doping and displacement fields shown to give rise to tunable flat bands \cite{gonzalez2017electrically, wolf2019electrically, cao2020tunable}. An important next step is to explore whether it is possible to engineer even flatter bands and enhance the zero-energy DOS, or extend the range of twist angles hosting flat bands, thereby further stabilizing and enriching the correlated phases that emerge in TBG.

A viable alternative proposed for realizing flat bands is provided by strained or buckled monolayer graphene. Here, structural deformations produce spatially varying strain fields that suppress the kinetic energy, thereby providing a route to flat band physics in a twist-free way \cite{guinea2010energy, levy2010strain}. Scanning tunneling microscopy/spectroscopy (STM/STS) measurements on buckled graphene superlattices have already observed the emergence of such flat bands \cite{mao2020evidence} and have also been predicted to be continuously tunable with buckling strength \cite{milovanovic2020band}. By depositing graphene on hBN or $\text{NbSe}_2$ \cite{mao2020evidence}, such buckled graphene superlattices are easily realizable and have been projected as ideal candidates for studying emergent electronic orders \cite{manesco2020correlations} and valley-dependent topology \cite{manesco2021correlation}. Given that buckling in monolayer graphene can give rise to flat bands with high DOS, it is natural to consider the interplay of buckling and moir\'{e} physics in TBG.

Recent theoretical and experimental works have shown that free-standing TBG at small twist angles undergoes spontaneous corrugation \cite{jain2016structure, rakib2022corrugation, wang2023bending, wang2024universal, soares2025anisotropic}. Here, the twist-induced moir\'{e} patterns generate intrinsic shear stresses, which, under relaxation, cause the bilayer to undergo periodic buckling \cite{wang2024universal}. Likewise, the different stacking areas for all hexagonal van der Waals materials give rise to in-plane strain fields \cite{kazmierczak2021strain}, which leads to in-plane relaxation \cite{nam2017lattice}. These periodic deformations, involving both in-plane relaxation and out-of-plane corrugations, significantly affect the flat bands of TBG, increasing their bandwidth and modifying their separation from the valence and conduction bands \cite{lucignano2019crucial, ho2021hall, rakib2022corrugation}. In terms of external in-plane strain, homo/heterostrain has also already been extensively studied, and it has been shown that this can give rise to even more flat bands \cite{qiao2018twisted, huder2018electronic, zhang2020correlation, li2025strain}. However, despite these observations, the relationship between buckling in TBG, beyond natural corrugations and simpler strains, and the moir\'{e}-driven flat band physics remains largely unexplored.

In this work, we go beyond the current understanding by combining externally controlled periodic buckling and moir\'{e} physics to study their combined effect on the electronic structure of TBG using atom-resolved modeling.
In particular, we consider TBG at both small and large twist angles, subject to a superlattice of buckles that has been shown to give rise to tunable flat bands in monolayer graphene \cite{mao2020evidence, milovanovic2020band}, thereby enhancing the flat band character of TBG. Experimentally, such buckled superlattices may be realized by depositing TBG on designed substrates \cite{guinea2010energy, phong2022boundary}. 

At large twist angles, we demonstrate that TBG first exhibits greater band flattening than monolayer graphene for similar buckling, but eventually yields more dispersive bands for large enough buckles. By analyzing the symmetries of AA- and AB-stacked graphene in comparison to TBG, we find that the combined effects of buckling and twisting break the inversion symmetry that protects the Dirac-point degeneracy, thereby opening a gap at the Dirac points. This gap opening provides a general mechanism for band flattening around zero energy for larger angles and small to moderate buckling. 

At the magic angle, we investigate the interplay between moiré-driven flat bands and buckling in TBG. In pristine magic-angle TBG, the flat bands exhibit a sublattice-symmetric charge distribution \cite{trambly2010localization, baldo2023defect}. Buckling, however, couples asymmetrically to the two valleys in graphene through strain-induced gauge fields \cite{vozmediano2010gauge}, which inherently breaks the sublattice symmetry. As a consequence, the two flattening mechanisms, twist-induced moiré patterns and periodic buckling, cannot be additive, but instead compete more with one another at the magic angle. Therefore, the band flattening at large angles induced by buckling does not persist down to the magic angle, as the magic-angle moir\'{e} bands are already very narrow, but instead, buckling there leads to more dispersive bands.

We additionally quantify the emergence of buckling-induced flat bands to probe their potential as a host of diverse correlated orders for non-magic angles. We do this across a wide range of twist angles by using the integrated density of states (IDOS) near zero-energy as a viable measure. Our results show that periodic buckling, with its strength serving as a tuning parameter, can, for a range of twist angles, more than double the IDOS compared to pristine TBG. Remarkably, near, but not at, the magic angle, we show that buckled TBG can rival pristine magic-angle TBG in its ability to support correlated phenomena. Interestingly, this establishes an intrinsic robustness of the flat bands of buckled TBG against twist-angle disorder.

The remainder of this work is organized as follows. We describe our model for TBG and the physics of periodic buckling in \cref{sec:level2}. In \cref{sec:level3} we investigate the effect of buckling on TBG at large twist angles, and show that a broken inversion symmetry leads to band flattening. In \cref{sec:level4}, we study the effect of buckling at the TBG magic angle on the electronic structure. We finally consider intermediate twist angles in \cref{sec:level5} and quantify the ability of buckled TBG to host flat bands using buckling strength as a tuning parameter. Finally, in \cref{sec:level6}, we summarize our results.

\section{\label{sec:level2} Modeling buckling in twisted bilayer graphene}
\subsection{Twisted bilayer graphene}
To capture the relevant physics, we describe TBG using the tight-binding Hamiltonian  
\begin{equation}
    \label{eq:hamiltonian}
    \mathcal{H} = -\sum_{i,j}t(|\mathbf{r}_i-\mathbf{r}_j|)c_i^{\dagger}c_j,
\end{equation}
where the hopping strengths $t$ are given by a distance-dependent Slater-Koster parametrization \cite{trambly2010localization, moon2013optical, lothman2022nematic, baldo2023defect}
\begin{equation}
\label{eq:sk}
\begin{aligned}
    t(|\mathbf{r}_i-\mathbf{r}_j|) &= t_0\,e^{\left(a_c-r_{ij}\right)/\lambda} \left(1-\left(\hat{\mathbf{r}}_{ij}\cdot\hat{\mathbf{z}}\right)^2\right) \\&+ t_{\perp} \, e^{\left(d_0-r_{ij}\right)/\lambda} \left(\hat{\mathbf{r}}_{ij}\cdot\hat{\mathbf{z}}\right)^2.
    \end{aligned}
\end{equation}
Here, $t_0 = 2.7\,\,\text{eV}$ and $t_{\perp} = 0.48\,\,\text{eV}$ are the in-plane and out-of-plane hopping strengths, respectively, and $\mathbf{r}_{ij} = \mathbf{r}_i-\mathbf{r}_j$. Furthermore, $a_c = 1.42\, \text{\AA}$ is the intralayer carbon-carbon distance, $d_0 = 3.35\, \text{\AA}$ is the equilibrium interlayer distance, and $\lambda = 0.184 \,a_c$ sets the decay length of the orbital overlap \cite{moon2013optical, lothman2022nematic, baldo2023defect}. We omit spin degrees of freedom as we only consider spin-degenerate states. We restrict the in-plane hopping to nearest-neighbors only, while for the interlayer coupling, we include interactions up to sixth-order nearest-neighbors. This has been shown to reproduce the correct band structure for magic-angle TBG, including the emergence of flat bands and a finite band gap separating the moir\'{e} bands \cite{lothman2022nematic, baldo2023defect}, and it also ensures compatibility with the selected buckling profile.  Truncations of the interlayer coupling at shorter distances fail to reproduce the correct band structure, while higher-order terms have minimal effect on the band structure of pristine or buckled TBG. Monolayer graphene can be modeled using the same geometry and Hamiltonian but with no interlayer coupling ($t_\perp$=0).

\begin{figure}[]
    \includegraphics[width=1\columnwidth]{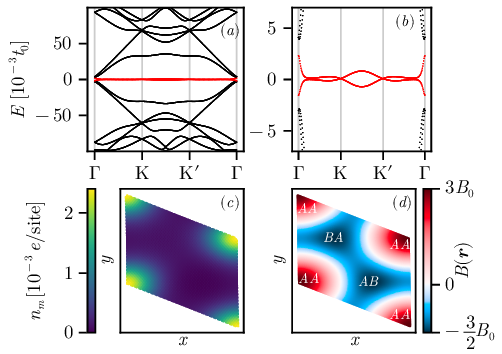}
    \caption{Band structure of magic-angle TBG (a), with zoom-in on moir\'{e} bands highlighted in red (b), charge density of the moir\'{e} bands (c), and the buckling-induced PMF of Eq.~\eqref{eq:tripmf} projected over the moir\'{e} unit cell (d). Indicated in (d) are the different stacking regions of TBG.}
    \label{fig:pmf}
\end{figure}

We define a commensurate moir\'{e} unit cell characterized by two integers $p$ and $q$, for the Hamiltonian Eq.~\eqref{eq:hamiltonian}, with the twist angle between the two graphene layers given by $\cos\theta = (3q^2-p^2)/(3q^2+p^2)$ \cite{shallcross2010electronic}. The twist angle $\theta$ is $0\degree$ $\left((p,q) = (0, 1)\right)$ for AA-stacked graphene and $60\degree$ $\left((p,q) = (1, 1)\right)$ for AB-stacked (Bernal-stacked) graphene. For all other angles studied in this work, we fix $p=1$ and vary $q$. Furthermore, we assume $q$ to be odd, as that yields smaller unit cells and is hence computationally more efficient. The size of the moir\'{e} unit cell is given by the moir\'{e} length $L_m = a/(2\sin(\theta/2))$, where $a = \sqrt3 \,a_c$ is the lattice constant of graphene \cite{shallcross2010electronic}. At the so-called magic angle, here $\theta \approx 1.20\degree$ generated by $(p, q) = (1, 55)$, the unit cell contains 9076 atoms and has a length $L_m \approx 47.6\,a$. We further neglect atomic relaxation effects, which are known to modify the local stacking configuration by shrinking the AA-stacked regions. While such relaxations renormalize the magic angle and enhance band gap opening \cite{nam2017lattice, si2016strain, ezzi2024analytical, kang2025analytical}, here, we focus on unrelaxed structures as a useful first approximation. Moreover, since we induce buckling actively through external means, it further justifies neglecting minor intrinsic relaxation effects. This means that the relative areas of the stacking domains remain unchanged, and the interlayer distance is kept constant. 
At the magic angle, the moir\'{e} pattern generates the characteristic flat band structure of TBG~\cite{trambly2010localization,andrei2020graphene}, as illustrated in \cref{fig:pmf}(a,b) 

We obtain the eigenenergies $E_{n, \mathbf{k}}$ and eigenvectors $\psi_{n, \mathbf{k}}$, with band index $n$ and momentum $\mathbf{k}$, of the spectrum through sparse diagonalization of the Hamiltonian Eq.~\eqref{eq:hamiltonian} and focus on the low energy region. We identify the four lowest-energy bands as the moir\'{e} bands, indicated in red in \cref{fig:pmf}(a,b). We denote the bandwidth of the moir\'{e} bands as $\Delta E_m$, chosen so that it captures the four moir\'{e} bands, but its value is not necessarily the same for every buckling configuration. We further resolve the moir\'{e} bands in real space by computing the site-resolved moir\'{e} charge density, defined as
\begin{equation}
    \begin{aligned}
        n_m(\mathbf{x}, \Delta E_m) &= e\int_{\Delta E_m}\rho(\mathbf{x},E)\,dE\\
        &=e\sum_{E_{n,\mathbf{k}}\in\Delta E_m} \left|\psi_{n,\mathbf{k}}\right|^2,
    \end{aligned}
\end{equation}
where $e$ is the elementary charge. Here, $\rho$ denotes the density of states (DOS). To obtain a smooth function, the discrete energies are broadened using a Lorentzian, mimicking finite-temperature effects. Computing $n_m(\mathbf{x}, \Delta E_m)$ for the four moir\'{e} bands, we see that they are primarily localized in the AA-stacked region, see \cref{fig:pmf}(c), with electrons equally divided among both sublattices \cite{trambly2010localization}. 
Closely related to the moir\'{e} charge density is the integrated DOS (IDOS), given by 
\begin{equation}
    \label{eq:idos}
    \text{IDOS}(\Delta E_U) = \int_{\mathcal{M}} n(\mathbf{x},\Delta E_U)\,d\mathbf{x}.
\end{equation}
The spatial integral sums up the contribution of the charge density $n$ of every site in the moir\'{e} unit cell $\mathcal{M}$ and thus the IDOS quantifies the available number of electronic states within an energy window $\Delta E_U$. Here, we allow for $\Delta E_U$ not being the same as the moir\'{e} bandwidth $\Delta E_m$ and instead consider it to correspond to the Coulomb interaction strength projected onto the moir\'{e} bands. In this way, the IDOS becomes a measure of the effective flatness seen by the interactions in the system, thus quantifying the propensity for developing correlated phenomena, such as superconductivity and magnetism \cite{yankowitz2019tuning, gonzalez2017electrically, wolf2019electrically}. 
For TBG, this relevant energy scale $\epsilon_U$ is typically taken to lie between 20 and 30 meV \cite{bernevig2021twisted, liu2025nodal}. Here, we set $|\Delta E_U| = 27$ meV, which, for simplicity, corresponds to the interval $\Delta E_U =\left[-5,5\right]\times10^{-3}t_0$. We also check intervals corresponding to the extremal projected Coulomb interaction strengths (20 and 30 meV) and find that the qualitative results remain unchanged. We note that there exist other measures of flatness than the IDOS, such as bandwidth or the effective mass of a band. While they are often adequate for well-separated, flat bands, here they fall short in capturing the contribution from multiple bands that are close in energy.

\subsection{Buckling}
Strain and buckling in monolayer graphene provide a powerful means to modulate the electronic properties \cite{si2016strain, hou2024strain}. Such structural deformations alter the equilibrium atomic positions, thereby modifying the hopping strengths between atoms. In monolayer graphene, to lowest order, this breaks the symmetry of the three equivalent nearest-neighbor hoppings as $t_0\mapsto t_i = t_0+\delta t_i$, with $i \in (1, 2, 3)$, for the three nearest neighbors. A low-energy description of this transformation is obtained when it is plugged back into Eq.~\eqref{eq:hamiltonian}. The result is an effective pseudomagnetic vector potential $\mathbf{A}$ that couples to the electronic degrees of freedom, given by \cite{vozmediano2010gauge, manesco2020correlations, manesco2021correlation}
\begin{equation}
    \mathbf{A} = \left(A_x,\,A_y\right) = \frac{1}{2ev_F} \left(\sqrt{3}\,(t_3-t_2),\,t_2+t_3-2t_1\right),
    \label{eq:vecpot}
\end{equation}
with $v_F = \frac{3}{2}t_0a_c$ denoting the Fermi velocity in graphene. Similar to conventional gauge theories, the vector potential has an associated field given by $\mathbf{B}_{\text{PMF}}=\nabla\times \mathbf{A}$, known as the strain-induced pseudomagnetic field (PMF). Although it behaves like a magnetic field and, under the right conditions, even allows for (pseudo) Landau quantization of the electronic states with characteristic cyclotron orbits \cite{milovanovic2020band, guinea2010energy}, the PMF is fundamentally different from a real magnetic field. Importantly, real magnetic fields break time-reversal symmetry, whereas PMFs preserve it, but instead break valley (sublattice) symmetry \cite{vozmediano2010gauge, moldovan2013electronic}. In graphene, it has been demonstrated that strain-induced PMFs can reach higher strengths than are achievable using real magnetic fields, reaching strengths equivalent to 300\,T \cite{levy2010strain}.

In this work, we focus on a periodic buckling configuration that has been identified in monolayer graphene as particularly promising for the emergence of flat bands \cite{milovanovic2020band, manesco2020correlations, manesco2021correlation}. Notably, this buckling configuration can be continuously tuned, with the flatness of the bands increasing with the strength of PMF in monolayer graphene. Experimentally, this buckling pattern has been realized by depositing graphene on substrates such as hBN or $\text{NbSe}_2$, which naturally give rise to buckled superlattices \cite{mao2020evidence}. Alternatively, a corrugated substrate can be engineered and positioned beneath the graphene layer to achieve a similar effect \cite{phong2022boundary}. Here, we aim to combine the intrinsic moir\'{e} potential in TBG with the buckling-induced PMF generated by such substrate engineering and to study its effect on the electronic structure. We do not consider other buckles reported in earlier works  \cite{milovanovic2020band}, as they are unable to localize electrons and instead only scatter them, preventing the formation of flat bands, and thus making them unsuitable for our goal. 

As we consider both buckling and moir\'{e} patterning, we must consider buckling periodic with the moir\'{e} pattern to be able to achieve a numerically viable setup. 

The PMF, which is empirically derived \cite{mao2020evidence}, is given by
\begin{equation}
    \label{eq:tripmf}
    \mathbf{B}(\mathbf{r}) =  B_{0}\sum_{i}\cos(\mathbf{b}_i\cdot(\mathbf{r}-\mathbf{r}_0)),
\end{equation}
where $B_0$ is the strength of the induced PMF in Tesla, and the three different $\mathbf{b}_i$ correspond to the two reciprocal moir\'{e} unit cell lattice vectors, the linear combination of them. Furthermore, $\mathbf{r}_0$ corresponds to the center of the buckle, serving as an additional parameter that allows us to shift the buckle center in the moir\'{e} unit cell. Setting $\mathbf{r}_0=0$ centers the buckle in the middle of the AA-stacked region. In \cref{fig:pmf}(d), we illustrate such centered buckling in magic-angle TBG. The PMF attains its maximum in the center of the AA-stacked region, and its minima in the center of the AB/BA-stacked regions. It is important to note that the average induced PMF is zero. As a result, the electronic states arising from this buckling are Bloch bands rather than pseudo Landau levels (pLL) \cite{fujimoto2025higher}. Fundamentally, because the PMF varies rapidly, it cannot give rise to well-defined pLLs. For magnetic quantization effects to emerge, the PMF instead needs to remain constant on the length scale of the magnetic length \cite{guinea2010energy, milovanovic2020band}. Thus, we do not expect to see well-defined pLLs in this work.
Unless stated otherwise, we assume $\mathbf{r}_0=0$, but we also explicitly investigate the non-zero $\mathbf{r}_0$.

For the PMF in Eq.~\eqref{eq:tripmf}, the deformation fields that generate it are unknown. As a consequence, we cannot directly modify the atomic positions and plug these into the Slater-Koster formalism of Eq.~\eqref{eq:hamiltonian}. However, following Refs.~\cite{mao2020evidence, milovanovic2020band, manesco2020correlations, manesco2021correlation}, we can still compute the changes to the in-plane hopping strengths $\delta t_i$ by inserting the ansatz $\delta t_i = \alpha_i\sin(\mathbf{b}_i\cdot\mathbf{r})$ into Eq.~\eqref{eq:vecpot} and fixing the $\alpha_i$ to match Eq.~\eqref{eq:tripmf}, for details see \cref{sec:app1}. In the end, the resulting changes to the three different nearest-neighbor hoppings, setting $\hbar = 1$, are given by
\begin{equation}
    \label{eq:hopping}
    \delta t_i = \frac{-\sqrt{3}ev_F}{2\pi} L_mB_0\sin(\mathbf{b}_i\cdot(\mathbf{r}-\mathbf{r}_0)).
\end{equation}
This expression is overall consistent with earlier works \cite{manesco2020correlations, manesco2021correlation,mao2020evidence, milovanovic2020band}.

Importantly, the modulation of the hopping amplitudes scales with both $L_m$, which varies across the different TBG systems studied here, and $B_0$. Experimentally, the degree of lattice buckling controls bond length variations and, therefore, the magnitude of the hopping modulation. Since these modulations set the relevant energy scale in Eq.~\eqref{eq:hamiltonian}, keeping this scale fixed, or equivalently the shape of the buckle, requires that the product $L_m B_0$ remain constant. As a result, smaller systems, corresponding to TBG at large twist angles, demand significantly larger $B_0$ to produce comparable buckles. This scaling is physically intuitive, as a similar buckle in a larger moir\'{e} unit cell corresponds to a smoother buckling profile with lower curvature, and thus a weaker effective PMF \cite{vozmediano2010gauge}.

Further, we note that Eqs.~\eqref{eq:tripmf} and~\eqref{eq:hopping} are also directly applicable to monolayer graphene, because they only change the intralayer coupling. In TBG, the situation becomes more complex because buckling may also modulate the interlayer distance between the two graphene sheets, which is not captured by these two equations. To address this, we consider identical buckles in each layer, so that the effective interlayer distance remains constant, and the Slater-Koster parameterization in Eq.~\eqref{eq:hamiltonian} remains valid. For substrate-induced buckling, this should be a good first approximation.

Taken together, we employ Eq.~\eqref{eq:hopping} to introduce controlled periodic buckling in TBG given in Eq.~\eqref{eq:hamiltonian} from large twist angles, down to the magic angle. We aim to investigate how buckling interacts with the moir\'{e} physics of TBG, especially focusing on possibilities to flatten the low-energy bands.
To estimate that we are in the correct parameter ranges, we compute the maximum modulations to the in-plane hopping using Eq.~\eqref{eq:hopping}. The $\delta t_i$ are bounded (from below) by $-\sqrt{3}ev_FL_mB_0/2\pi$. At the magic angle, where $L_m \approx 47.6\,a$, and using a PMF strength $B_0=50$ T, this suggests $\delta t_i \gtrsim -0.14$ eV. This represents the largest effective buckling strength considered in this work. This is smaller in magnitude than previous experimental works \cite{mao2020evidence}, which have investigated buckling in monolayer graphene at $L_m \approx 14$ nm and $B_0 = 120$ T, corresponding to $\delta t_i \gtrsim -0.40$ eV.

\section{\label{sec:level3} Enhanced flattening and symmetry breaking in large-angle TBG}
We start by investigating the effect of periodic buckling, centered on the AA-stacked region, on the low-energy electronic structure of large-angle TBG, choosing a representative twist angle $\theta = 3.15 \degree$. We first show that buckled large-angle TBG exhibits enhanced band flattening compared to monolayer graphene. We also compare the effects of buckling in AA and AB-stacked bilayer graphene and thereby uncover the importance of broken inversion symmetry.

\begin{figure}[]
    \includegraphics[width=1.0\columnwidth]{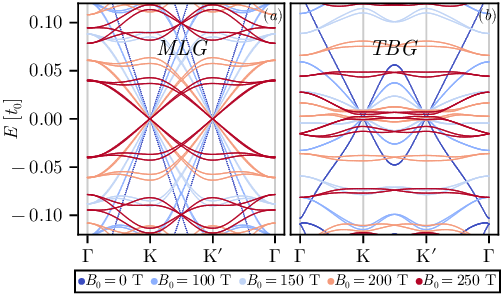}
    \caption{Low-energy band structure of buckled monolayer graphene (a) and large-angle TBG at $\theta = 3.15 \degree$ (b) for different PMF strengths, with the same moir\'{e} unit cell in (a,b).}
    \label{fig:interlayerenhance}
\end{figure}

In \cref{fig:interlayerenhance} we plot the band structures of pristine and buckled graphene (a) and large-angle TBG (b) at different PMF strengths $B_0$, using the same moir\'{e} unit cell for both systems.
Pristine monolayer graphene and large-angle TBG have comparable Fermi velocities and other low-energy electronic properties, a direct consequence of large-angle TBG's nearly homogenous charge density \cite{trambly2010localization}. However, significant differences emerge after buckling. In graphene, \cref{fig:interlayerenhance}(a), buckling continuously and monotonously flattens the Dirac cones, which has been attributed to electronic confinement due to the non-uniform strain \cite{milovanovic2020band}. This leads to a substantial reduction in the Fermi velocity, resulting in flatter bands. In large-angle TBG, there is an even stronger Fermi velocity reduction, resulting in an increased band flattening. We attribute this enhancement to the interlayer coupling in TBG. In its absence, electrons in the individual monolayers tend to localize in regions of high PMF, tied to the buckle. However, when interlayer coupling is introduced, electrons are not only able to tunnel in-plane, but also between layers, effectively decreasing the in-plane kinetic energy. This enhances the in-plane electronic confinement and localization caused by the PMF, overall resulting in a flatter dispersion. This mechanism, in fact, resembles the band-flattening behavior observed in pristine TBG as the twist angle decreases towards the magic angle. In the latter case, it is the increasing size of the AA-stacked regions causing the moir\'{e} potential to grow, which promotes greater electron localization. Moreover, unlike monolayer graphene, the band flattening in TBG is not fully monotonous with increasing PMF strength $B_0$. For TBG, buckling initially flattens the spectrum with increasing PMF strength, but, for example, for $B_0 = 250 $ T, the resulting four central bands are more dispersive compared to $B_0 = 200$ T.

Furthermore, analogous to monolayer graphene \cite{mao2020evidence, milovanovic2020band}, the continuum of bands away from the low-energy moir\'{e} bands in TBG transforms into a set of well-isolated bands at higher energies with increasing PMF strength. In fact, these higher-energy isolated flat bands are significantly flatter in large-angle TBG than in monolayer graphene. Unlike monolayer graphene, in TBG, these bands also sporadically hybridize with the lowest energy moir\'{e} bands. We find that these higher-energy isolated flat bands emerge at all twist angles. They may be ideal candidates for hosting fractional Chern insulating states, as has been recently postulated in buckled AB-stacked graphene \cite{fujimoto2025higher}. We note that due to the spatially strongly varying PMF, these flat bands do not correspond to pLLs.

\begin{figure}[]
    \includegraphics[width = 1.0\columnwidth]{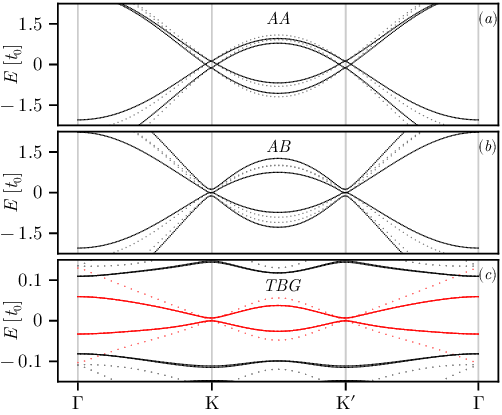}
        \label{fig:3a}
        \label{fig:3b}
        \label{fig:3c}
    \caption{Low-energy band structure for pristine and buckled AA-stacked bilayer graphene with $B_0=10000$ T and $L_m= a$ (a), AB-stacked bilayer graphene with $B_0= 10000$ T and $L_m= a$ (b), and large-angle TBG at $\theta = 3.15\degree$, corresponding to $L_m\approx18.2\,a$, with $B_0= 100$ T (c) to keep that the product $L_m B_0$ roughly constant. Red bands are the moir\'{e} bands. Dotted bands correspond to pristine, non-buckled bands. Note in (c) the smaller scale on the y-axis.}
    \label{fig:symmbreak}
\end{figure}

Additionally, we find in \cref{fig:interlayerenhance} that buckling large-angle TBG gaps the spectrum at the Dirac points, $K, K^{'}$, which does not occur in pristine TBG or in monolayer graphene. To highlight this, we compare large-angle TBG to AA and AB-stacked bilayer graphene. Due to the small size of the unit cell in the latter two systems, we need to employ a strong PMF to reach effective lattice changes. For a fair comparison, where we take large-angle TBG at $B_0 = 100$\,T as a reference, we can adjust the corresponding $B_0$ in bilayer graphene such that the product $L_mB_0$ is roughly constant. This, however, yields little buckled-induced change in the electronic structure of the low-energy band structure of either AA- or AB-stacked bilayer graphene. Instead, to induce noticeable changes and highlight the gap features, we apply a $B_0$ to AA- and AB-stacked graphene that is two and a half times stronger. We plot the resulting band structures of the pristine and buckled bilayers in \cref{fig:symmbreak}. In both monolayer graphene \cref{fig:interlayerenhance}(a) and the untwisted bilayers \cref{fig:symmbreak}(a,b), the Dirac points remain protected by combined $\mathcal{C}_{2z}\mathcal{T}$ symmetry \cite{castro2009electronic}, where the $z$-axis is defined perpendicular to the two graphene sheets. The induced PMF couples with opposite sign to the two valleys in graphene. This breaks the sublattice symmetry and induces a spatially inhomogeneous mass term. However, this inhomogeneous mass term alone is not sufficient to lift the degeneracy at the Dirac point, since it does not break the $\mathcal{C}_{2z}$ inversion symmetry. Also, as mentioned in \cref{sec:level2}, buckling never breaks $\mathcal{T}$. Thus, the degeneracy at the Dirac points is preserved for monolayer, and AA- and AB-stacked bilayer graphene. In contrast, in \cref{fig:symmbreak}(c), the Dirac points are gapped with buckling for the twisted bilayers. Specifically, the slight layer misalignment in TBG, together with buckling, is enough to break inversion symmetry, and thus gap the spectrum \cite{po2018origin}. This opening of a gap at the Dirac point is interesting on a fundamental level, but importantly, also has the consequence of both effectively flattening the four moir\'{e} bands. In fact, at large twist angles, the bandwidth at the center of the Brillouin zone $\Gamma$, where it reaches its maximum, is sufficiently large so that the emergence of the gap at the Dirac point enhances the overall flatness. This resembles closely the alignment or encapsulation of TBG with hBN \cite{long2023electronic}. We further note that the size of the gap does not grow monotonically with increasing strength of the PMF, something that would happen in the presence of a global sublattice-symmetry-breaking mass term. We attribute this to the non-local nature of the inhomogeneous mass term.

We also investigate the role of the buckling center position ${\bf r}_0$ in large-angle TBG. However, due to the nearly uniform charge density in large-angle TBG, similar to that of monolayer graphene, we find that the position of the buckling center has only a minimal impact on the electronic structure. Aside from an occasional shift in energies, the qualitative features remain the same.

We conclude that large-angle TBG exhibits enhanced band flattening compared to monolayer graphene due to the presence of interlayer coupling. We interpret this as an enhancement of electronic localization in regions of strong PMF. Also, unlike monolayer graphene, the band flattening in TBG is not monotonous. Moreover, buckling in TBG lifts the degeneracy at the Dirac points by breaking the $\mathcal{C}_{2z}$ inversion symmetry, leading to flatter bands. In monolayer graphene and AA- and AB-stacked bilayer graphene, the degeneracy at the Dirac point remains.

\section{\label{sec:level4} Competing mechanisms at the magic angle}
\begin{figure}[] 
    \includegraphics{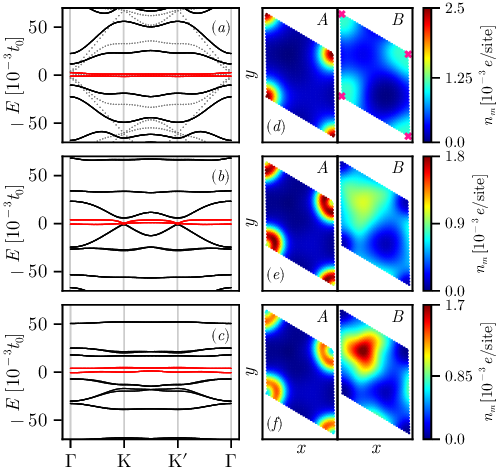}
    \caption{Low-energy band structures (a-c) and corresponding top layer sublattice-resolved (A, B) moir\'{e} charge densities $n_m$ (d-f) for buckled magic-angle TBG with buckle centered in the AA-stacked region (${\bf r}_0 = 0$), indicated with pink crosses in (d), and buckling strengths $B_0 = 10$\,T (a,d), 25\,T (b,e), and 35\,T (c,f). Red bands are the four energy bands that most resemble the moir\'{e} bands, gray bands in (a) are those of pristine magic-angle TBG.}
    \label{fig:AAmagicband}
\end{figure}

After discussing buckling in large-angle TBG, we turn to the magic angle, where TBG exhibits moir\'{e}-like physics rather than graphene-like physics. We buckle magic-angle TBG at multiple PMF strengths and analyze the resulting band structures and charge densities in \cref{fig:AAmagicband}. We first choose to center the buckle in the AA-stacked region ${\bf r}_0=0$, with the idea to enhance the electron confinement that is already there from twisting, but later also center the buckle in the AB-stacked region, which, at the magic angle, has a negligible charge density \cite{trambly2010localization}. 

To investigate the effect of buckling on the electronic structure of magic-angle TBG, we plot in \cref{fig:AAmagicband} both the band structures and (a-c) and sublattice-resolved moir\'{e}-charge densities (d-f) for increasing PMF strength. The charge densities are visualized in the top layer only, but are the same for the bottom layer. 
The most striking feature is the emergence of a sublattice charge polarization, causing a sublattice asymmetry in \cref{fig:AAmagicband}(d-f). Already in \cref{fig:AAmagicband}(d) where the band structure is not so affected by the buckling (compare gray and black lines in \cref{fig:AAmagicband}(a)) there is an enhancement of charge localized around the AA-stacked region on one sublattice (A), while there is a depletion for the other sublattice (B). Such sublattice polarization is also known to occur in monolayer graphene with the same buckling profile \cite{mao2020evidence, milovanovic2020band}. It will also occur for other periodic buckling patterns because of how the PMF generally couples differently to the two valleys in graphene. In contrast to pristine magic-angle TBG, which has a sublattice symmetric charge density, buckling explicitly and inhomogeneously polarizes the sublattices. Hence, because of the difference in where these two flattening mechanisms want to localize the electrons, they become competing and are thus fundamentally incompatible with one another. As the buckling strength is further increased in \cref{fig:AAmagicband}(e,f), the sublattice polarization becomes more pronounced, but it is not enough to create isolated islands of electrons. It is worth mentioning that these electrons that are localized by the PMF do not correspond to pLLs, but are instead confined to regions with a strong PMF \cite{milovanovic2020band}. As a consequence of the pronounced sublattice polarization in buckled TBG, describing the system in terms of moir\'{e} physics becomes unfitting. 

Focusing on the low-energy moir\'{e} band structure, we find that the buckling-induced PMF does not permit continuous flattening of the spectrum, either at the magic angle as seen in \cref{fig:AAmagicband}(a–c), or at large twist angles, as discussed in \cref{fig:interlayerenhance}(b), in contrast to monolayer graphene. 
Moreover, we find that the gap induced at the Dirac point, due to the breaking of the inversion symmetry, causes substantial changes to the moir\'{e} bands with buckling.
Already at low to moderate PMF strengths in \cref{fig:AAmagicband}(a), the gap is clearly noticeable and, in contrast to large-angle TBG, the original bandwidth at $\Gamma$ is now sufficiently narrow so that the gap instead increases the overall bandwidth, yielding a more dispersive bandstructure, compared to pristine TBG. Interestingly, for moderate PMF strength \cref{fig:AAmagicband}(b), the conduction and valence bands even hybridize with the original four moir\'{e} bands. This roughly mimics the dispersion of large-angle buckled MLG, except for the four extra moir\'{e} bands in the middle.  

When it comes to the continuum of bands away from the moir\'{e} bands, they gradually evolve into a set of isolated bands with increasing PMF strength as illustrated through \cref{fig:AAmagicband}(a-c). As the PMF strength increases, these bands can, unlike graphene, occasionally hybridize with each other or the four original moir\'{e} bands, see e.g.~\cref{fig:AAmagicband}(b). For strong PMF, where buckling is the dominant flattening mechanism, these isolated bands can become extremely flat, as shown in \cref{fig:AAmagicband}(c). We compare the bandwidth of these emergent flat bands to those of monolayer graphene with the same buckled moir\'{e} unit cell and PMF strengths (see \cref{sec:app2} for details). We find that the TBG bandwidths are significantly smaller, often reduced by up to a factor of 2-3. As an example, at a large $B_0=35$ T, we find that the bandwidth of the first isolated band above the Fermi energy is about 70 \% smaller than in monolayer graphene. 

\begin{figure}[]
    \includegraphics{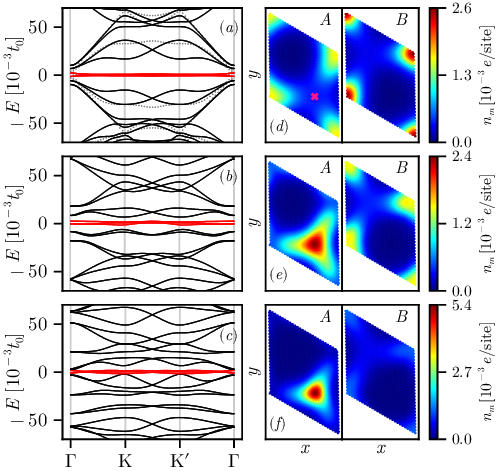}
    \caption{Same as \cref{fig:AAmagicband}, but with buckle centered on the AB-stacked region, indicated with pink cross in (d).}
    \label{fig:ABmagicband}
\end{figure}

Finally, we center the buckle in the AB-stacked region to investigate the importance of the location of the buckle. In \cref{fig:ABmagicband} we again plot the band structure and charge density for the same buckling as in \cref{fig:AAmagicband}. Here, because of the AB-stacking at the center of the buckle, there is an asymmetry between the charge accumulated in the top and bottom layers. We focus on the top layer, since it turns out to be more susceptible to changes in the PMF strength. When we compute the charge density in the bottom layer, it initially looks similar to the top layer. For moderate to strong buckles, however, electrons are mostly localized in the top layer, leaving almost no charge in the bottom layer. We attribute this layer dependence to the interplay between buckled-induced sublattice polarization and the different local stacking configurations in both layers.

For weak buckles in \cref{fig:ABmagicband}(a), we obtain a noticeable difference in the spectrum compared to \cref{fig:AAmagicband}(a). Here, the continuum of bands away from the four moir\'{e} bands initially becomes denser in energy, with more bands appearing in the same energy window, instead of transforming into a series of isolated bands. This continuum of bands is robust up to moderate PMF strengths, as illustrated in \cref{fig:ABmagicband}(b), but for large enough PMF strengths, shown in \cref{fig:ABmagicband}(c), it behaves similarly to the buckle centered in the AA-stacked region, where instead isolated bands start to emerge, although with larger bandwidth compared to \cref{fig:AAmagicband}(c). Unlike for the buckle centered in the AA-stacked region, the bands are, however, still not well-isolated, and instead, the whole spectrum is more dispersive. This overall behavior can readily be explained by the fact that in pristine magic-angle TBG, electrons are primarily localized in the AA-stacked region, see \cref{fig:pmf}(c). 
Consequently, a considerable perturbation in the AB-stacked region is required to induce noticeable changes in the electronic structure. This is similar to the behavior of atomic defects that also require a larger strength in the AB-region to evoke a noticeable change \cite{baldo2023defect}. Still, several similarities to the AA-centered buckles remain. First, as shown by the sublattice-polarized charge densities in \cref{fig:ABmagicband}(d-f), there is still competition between electronic localization driven by twisting and buckling. Second, inversion symmetry breaking persists, leading to a gap at the Dirac points. However, a key difference between \cref{fig:ABmagicband}(c) and \cref{fig:AAmagicband}(c) lies in the spatial localization of the electrons. For AB-region buckles, electrons are confined to the center of the buckle, whereas for AA-region buckles, they form a ring around the center in one sublattice (A), as also known to occur in monolayer graphene \cite{mao2020evidence, milovanovic2020band}, and are also localized on the BA-stacked region in the other sublattice (B).
 
We conclude that the flattening of the moir\'{e} bands due to twisting at the magic-angle in TBG and flattening due to the PMF fundamentally compete with each other since the two flattening mechanisms localize electrons in different regions on different sublattices, making them fundamentally incompatible. This incompatibility persists regardless of the buckle's location. The main distinction between AA- and AB-buckled TBG is that the former exhibits flatter isolated bands, while the latter features a denser energy spectrum.

\section{\label{sec:level5} Enhanced band flattening close to the magic angle}
After exploring both the large- and magic-angle limits of TBG, we finally focus on the intermediate regime near the magic angle, studying twist angles from the magic angle up to around $2\degree$. In \cref{sec:level3} and \cref{sec:level4}, we uncovered that buckling large- and magic-angle TBG can induce additional flatness into the low-energy bands. We are here interested in how suitable TBG is as a platform for hosting buckling-induced flat bands, as quantified by the IDOS introduced in \cref{sec:level2}. 
This would open up the possibility of inducing electronic ordering, such as superconductivity and magnetism, also beyond the magic angle. Here, we focus on the low-energy regime around zero energy, using $\Delta E_U$ as the relevant energy regime to probe the IDOS relevant for correlation effects, but note that additional gating of TBG can also open for using higher-energy bands. However, while higher energy bands can become well-isolated and flat for large buckling, they generally only consist of two spin-degenerate bands, while there are four spin-degenerate  (buckled deformed) moir\'{e} bands, thus providing more IDOS near zero energy.  We introduce buckles in the center of the AA-stacked regions, since they give rise to flatter structures sooner than buckles centered in the AB-stacked region, see \cref{sec:level4}. 
From Eq.~\eqref{eq:hopping}, we see that both the size of the moir\'{e} unit cell $L_m$ and the PMF strength $B_0$ dictate the magnitude of the modulations to the in-plane hopping. We thus hold the product $L_mB_0$ constant across all twist angles, ensuring a fair comparison of the effect of buckling on the low-energy electronic structure. Furthermore, we use values of $B_0$ from 0 T to 50 T at the magic angle as the range of strengths at other twist angles, and denote these by $B_M$, with corresponding moir\'{e} length $L_M$. The corresponding $B_0$ at other twist angles is related through $B_0$ = $B_M L_M/L_m$.

\begin{figure}[]
    \includegraphics{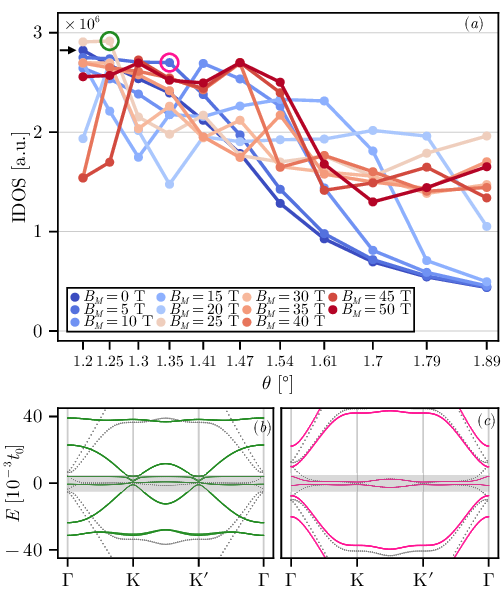}
    \caption{IDOS as a function of twist angle $\theta$ (a) and prototypical bandstructures (b,c), marked by matching colored circles in (a). Arrow indicates the IDOS for pristine magic-angle TBG. Light gray bands in (b) and (c) are band structures of pristine TBG at same twist angles. Dark gray boxes represent integration window set by $\Delta E_U$.}
    \label{fig:phasespace}
\end{figure}

We compute the IDOS for $\Delta E_U$ as a function of twist angle and buckling strength and compactly present the results in \cref{fig:phasespace}(a). For twist angles away from the magic angle, we see that buckling can significantly and consistently enhance the flatness of the spectrum. Here, the presence of buckles can more than double the IDOS of pristine TBG. For low $B_M$, the enhanced flatness is primarily driven by the gap opening at the Dirac points due to inversion symmetry breaking, as illustrated by a prototypical low-energy band structure in \cref{fig:phasespace}(c) extracted at pink circle in (a), and also discussed in \cref{sec:level3}. At moderate to large $B_M$ and near the magic angle, an increased IDOS can be the result of hybridization between buckling-induced emergent flat bands and the moiré bands, consistent with our observations in \cref{sec:level4} and illustrated with the band structure in \cref{fig:phasespace}(b) extracted at green circle in (a).
However, while smaller twist angles progressively flatten the Dirac cones, buckling can cause the moir\'{e} bands to flatten less or even become more dispersive, and hence lead to a smaller IDOS than the pristine case. Overall, this is due to the emergent competition between the two flattening mechanisms: the strong sublattice polarization induced by buckling negates the effect of twist-induced flattening.
As a consequence, buckling does not monotonically increase the IDOS near the magic angle, implying that no clear trends of IDOS as a function of buckling strength exist. However, for slightly larger angles we do find that buckling becomes the dominant flattening mechanism, and in turn can transform the moir\'{e} bands into very flat bands, as observed in \cref{fig:AAmagicband}(c). 

We conclude that buckling generally enhances band flatness at twist angles larger than the magic angle, compared to pristine TBG at the same angles. Near the magic angle, buckled TBG remains comparable to pristine magic-angle TBG in its ability to host flat bands. As a consequence, buckled TBG becomes much more robust to twist-angle disorder.

\section{\label{sec:level6} Conclusions}
In this work, we show how active buckling of TBG, achievable using, for example, substrate engineering, influences its low-energy electronic structure across a wide range of twist angles. Our work extends previous studies on buckled monolayer graphene as a platform for strongly correlated electronic phases by introducing buckling into TBG. TBG itself is already an excellent candidate for realizing such phases at magic twist angles due to flat moir\'{e} bands emerging from twisting.

In the large-angle limit, we find that TBG's moir\'{e} bands can be continuously flattened by increasing buckling strength. This behavior is similar to that of monolayer graphene, but with two key differences. First, the combination of twist and buckle breaks the symmetry that protects the Dirac points, leading to the opening of a gap, which generates a slightly flatter spectrum. Second, the interlayer coupling in TBG allows for a greater degree of band flattening than is achieved in monolayer graphene for the same PMF strengths. Overall, we find that buckling can significantly enhance large-angle TBG's capacity to form flat bands. For example, for twist angles close to $2\degree$, the integrated density of states (IDOS) in an energy range set by the effective Coulomb interaction projected onto the magic-angle flat bands can more than double for specific buckling strengths.

Close to the magic angle, we find that twist-induced band flattening and buckling-induced band flattening compete in how they localize electronic states. Specifically, buckling breaks sublattice symmetry, but is instead preferred by twist-induced flattening. Buckling thus induces a sublattice polarization, which is absent in the pristine moir\'{e} bands, making the two mechanisms fundamentally incompatible. Moreover, compared to large-angle TBG, the initial bandwidth near $\Gamma$ is now too narrow so that the gap generated by the inversion symmetry breaking at the Dirac points instead causes the bands to become overall more dispersive. Nonetheless, like in monolayer graphene, buckling in TBG can still develop a series of well-isolated bands at higher energy, and now with bandwidths often reduced by a factor of 2-3 compared to monolayer graphene.
Moreover, in monolayer graphene, these bands remain well-isolated, while in TBG, they occasionally hybridize with the moir\'{e} bands, especially close to the magic angle. This hybridization of additional flat bands, together with the gap opening, leads to the low-energy IDOS approaching or even slightly exceeding that of pristine magic-angle TBG for a range of PMFs and twist angles.  

Our results establish buckled TBG as a versatile and robust platform for studying strongly correlated electronic phenomena, and can even enhance them due to a higher low-energy DOS. For example, buckling can easily counteract twist angle disorder by helping to keep a high flatness of the low-energy bands independent of the twist angle. Importantly, all effects considered here should be experimentally accessible. Previously, similar periodic buckling has been implemented in monolayer graphene \cite{mao2020evidence}, and strain-induced PMFs of several hundreds of Tesla have also been observed in monolayer graphene \cite{levy2010strain}, much higher than required to reach the effects of buckling studied in this work. 

\section*{\label{sec:level7} Data availability}
The data supporting this work, as well as the code to generate the figures, are openly available at Zenodo \cite{zenodo}.

\begin{acknowledgments}
We thank L.~Baldo and P.~Holmvall for fruitful discussions. This work was supported by the European Research Council (ERC) under the European Union’s Horizon 2020 research and innovation programme (ERC-2022-CoG, Grant agreement No.~101087096). Views and opinions expressed are, however, those of the author(s) only and do not necessarily reflect those of the European Union or the European Research Council Executive Agency. Neither the European Union nor the granting authority can be held responsible for them. The calculations were enabled by resources provided by the National Academic Infrastructure for Supercomputing in Sweden (NAISS), partially funded by the Swedish Research Council through grant agreement no.~2022-06725. 
\end{acknowledgments}

\appendix
\section{Derivation of hopping modulations \label{sec:app1}}
In \cref{sec:level2}, we present in Eq.~\eqref{eq:hopping} a relationship describing the modulations in in-plane hopping strength due to an applied buckle. Here, we derive this expression in more detail. We start with the ansatz $\delta t_i = \alpha_i\sin(\mathbf{b}_i\cdot\mathbf{r})$ \cite{manesco2020correlations,manesco2021correlation} and plug this into Eq.~\eqref{eq:vecpot}. The buckling-induced vector potential is related to the PMF as $\mathbf{B}_{\text{PMF}}=\partial_x\mathbf{A}_y-\partial_y\mathbf{A}_x$, which gives
\begin{equation}
    \begin{aligned}
    \label{eq:appB}
    \mathbf{B}_{\text{PMF}} &= \frac{1}{2ev_F} (\alpha_2b_{2,x}\cos(\mathbf{b}_2\cdot\mathbf{r}) + \alpha_3b_{3,x}\cos(\mathbf{b}_3\cdot\mathbf{r}) \\&-2\alpha_1b_{1,x}\cos(\mathbf{b}_1\cdot\mathbf{r}))
    +\frac{\sqrt{3}}{2ev_F}(\alpha_2b_{2,y}\cos(\mathbf{b}_2\cdot\mathbf{r}) \\&- \alpha_3b_{3,x}\cos(\mathbf{b}_2\cdot\mathbf{r})).
    \end{aligned}
\end{equation}
Next, we use the reciprocal moir\'{e} vectors
\begin{equation}
    \begin{aligned}
    \mathbf{b}_1 &=\frac{2\pi}{L_m}\left(\frac{1}{\sqrt{3}}, \,-1\right)\\
    \mathbf{b}_2 &=\frac{2\pi}{L_m}\left(\frac{-2}{\sqrt{3}}, \,0\right)\\
    \mathbf{b}_3 &= -\mathbf{b}_1-\mathbf{b}_2= \frac{2\pi}{L_m}\left(\frac{1}{\sqrt{3}}, \,1\right),
    \end{aligned}
    \label{eq:moirerecip}
\end{equation}
and fix the $\alpha_i$ to match with Eq.~\eqref{eq:tripmf}. Coincidentally, one finds that all the $\alpha_i$ are equal.  This leads to 
\begin{equation}
    \alpha_i=-\frac{\sqrt{3}ev_F}{2\pi}L_mB_0,
\end{equation}
which finally yields Eq.~\eqref{eq:hopping} in the main text.
We note that this expression differs by a factor of $1/2$ from Refs.~\cite{manesco2020correlations, manesco2021correlation} and yields comparable spectra of works that follow a different ansatz if $B_0$ is rescaled to $B_0/4$ \cite{mao2020evidence, milovanovic2020band}. These prefactors do not influence any conclusions.

\section{Monolayer graphene \label{sec:app2}}
\begin{figure}[]
    \includegraphics{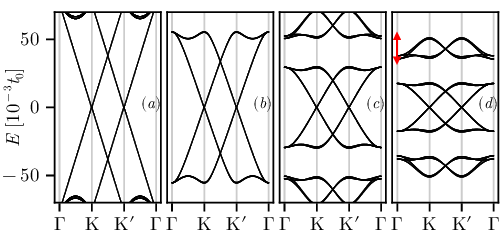}
    \caption{Low-energy band structure of pristine (a) and buckled (b-d) monolayer graphene calculated using the magic-angle TBG moiré unit cell for the buckling periodicity for $B_0 = 0$ T (a), $B_0 = 10$ T (b), $B_0 = 25$ T (c), and  $B_0 = 35$ T (d). Red arrow indicates the bandwidth of the isolated band discussed in the Appendix text.}
    \label{fig:graphene}
    \label{fig:appa}
    \label{fig:appb}
    \label{fig:appc}
    \label{fig:appd}
\end{figure}

In \cref{fig:AAmagicband}(a-c), we find that the bandwidths of both the Dirac cone and emergent isolated bands in buckled graphene are larger than buckled TBG. Here, we complement this by plotting the effect of buckling on monolayer graphene in \cref{fig:graphene}, which allows us to compare the various bandwidths directly. As an example, we find that the first set of well-isolated bands in buckled graphene at $B_0 = 35$ T has a bandwidth of around $16 \cdot 10^{-3}\,t_0$, see red double-arrow, which is significantly larger than in magic-angle TBG, with a bandwidth of around $5 \cdot 10^{-3}\,t_0$ for the same first set of higher-energy isolated bands. Furthermore, as discussed in \cref{sec:level2}, we find a continuous flattening of the Dirac cones, which has been observed in previous works \cite{mao2020evidence, milovanovic2020band}, and a preserved Dirac point degeneracy due to the absence of inversion-symmetry breaking.

\end{document}